\documentstyle[12pt]{article}
\begin{document}
\begin{flushright} 
NDA-FP-26\ \ \ \ \ \ \ \\
March 1996\ \ \ \ \ \ \ \\
\end{flushright}

\vfill

\begin{center}
{\large
\bf 
The Dirac Hamiltonian in an Aharonov-Bohm gauge field
and its self-adjoint extensions\\
}

\vfill

\renewcommand{\thefootnote}{\fnsymbol{footnote}}
{\large
Kazuhiko ODAKA\footnote[1]{e-mail address: odaka@cc.nda.ac.jp}~~and~~Kazuya  SATOH}

\vskip 5mm
{\it
Department of Mathematics and Physics

National Defence Academy, Yokosuka, 239, JAPAN}

\end{center}

\vfill

\begin{abstract}
By using the spherical coordinates in 3+1 dimensions we study the self-adjointness of the Dirac Hamiltonian in an Aharonov-Bohm gauge field of an infinitely thin magnetic flux tube. It is shown that the angular part of the Dirac Hamiltonian requires self-adjoint extensions as well as its radial one. The self-adjoint extensions of the angular part are parametrized by a 2 $\times$ 2 unitary matrix.

\vskip 3cm

\end{abstract}
\newpage
Aharonov and Bohm (AB) \cite{1} consider a charged particle scattered by a infinitely thin magnetic flux tube in order to clarify the significance of the vector potential in the quantum theory. The theoretical study of this phenomenon is related with many interesting and fundamental problems in quantum mechanics. One of them is to check the self-adjointness of the Hamiltonian. If it is not self-adjoint, we must construct its self-adjoint extensions \cite{2}. The customary condition, that the wave function is restricted to be nonsingular everywhere, leads to the result that the charged particle does not touch the magnetic flux. This result is true for the nonrelativistic spinless particle \cite{1}, while it is meaningless for the Dirac particle. Because the radial part of the Dirac Hamiltonian in the AB gauge field of the infinitely thin magnetic flux tube is not self-adjoint on such a condition \cite{3}. Some authors estabrish the  one-parameter family of its self-adjoint ex!
 tensions\footnote{Similar problems have been studied in monopole physics\cite{4}.} of which the wave function does not vanish at the origin \cite{5}. 

In the present paper, we analyze the Dirac Hamiltonian in the AB gauge field by using the spherical coordinates in 3+1 dimensions, since the use of them is inevitable in solving the eigenvalue equation for the Dirac Hamiltonian in the presence of a Coulomb and a magnetic monopole fields as well as the AB gauge field \cite{6,7}. 
It is shown that the angular part of the Hamiltonian requires the self-adjoint extensions similarly to the radial one discussed in \cite{5}.  The self-adjoint extensions of the angular part are parametrized by a  2$\times$2 unitary matrix. This situation remains in the case that the Coulomb and the magnetic monopole fields are introduced.  

The eigenvalue equation of the Dirac Hamiltonian $\hat H(F)$ in the AB gauge field $\vec A~=~F/(rsin\theta)\vec e_\phi$, where $F$ is the magnetic flux located at the z-axis, is
\begin{equation}
-i[ \alpha_r\frac{1}{r}(\frac{\partial}{\partial r}r)-
\alpha_r\beta\frac{1}{r}\hat K(F) +i\beta m]\psi(r,\theta,\phi)~=~ E\psi(r,\theta,\phi)
\end{equation} 
where
\begin{equation}
\hat K(F)~=~ \beta[\Sigma_\theta(i\frac{1}{sin\theta}(\frac{\partial}{\partial \phi}-iF)) + 
\Sigma_\phi(-i\frac{\partial}{\partial\theta}) + 1].
\end{equation}
We have made use of the notation~
$\alpha_r = \alpha_1 sin\theta cos\phi+
\alpha_2 sin\theta sin\phi+\alpha_3cos\theta,~$
$\Sigma_\theta = \Sigma_1cos\theta cos\phi+\Sigma_2cos\theta sin\phi-\Sigma_3 sin\theta~$ and 
$\Sigma_\phi = -\Sigma_1 sin\phi+ \Sigma_2 cos\phi~$ and our study will be carried out in the following representaion,
\begin{equation}
\alpha_i~=~\left[
\begin{array}{cc}
0           &\sigma_i\\
\sigma_i    &0
\end{array}
\right],~~~
\beta~=~\left[
\begin{array}{cc}
1 &0  \\
0 &-1  
\end{array}
\right],~~~
\Sigma_i~=~\left[
\begin{array}{cc}
\sigma_i &0  \\
0 & \sigma_i  
\end{array}
\right].
\end{equation}
Since the Hamitonian $\hat H(F)$ commutes with the operator $\hat K(F)$, we can choose simultaneous eigenfunctions of $\hat H(F)$ and $\hat K(F)$. It must be noted that the Hamiltonian must be self-adjoint therefore the operator $\hat K(F)$ is also required to be so. 
 
In order to derive the separation of variables form of the equation (1), we apply a similarity transformation\cite{6} given by
\begin{equation} 
S_0~=~\frac{1}{r(sin\theta)^{1/2}}exp[-i\frac{\phi}{2}\Sigma_3]exp[-i\frac{\theta}{2}\Sigma_2]
\end{equation}
to the wave function $\psi(r,\theta,\phi)$ and the Hamiltonian $\hat H(F)$,
\begin{equation}
\Psi (r,\theta ,\psi)~=~S^{-1}_0 \psi (r,\theta,\phi),
\end{equation}
\begin{equation}
S^{-1}_0 \hat H(F) S_0~=~\hat h(F)~=~
-i[\alpha_3\frac{\partial}{\partial r}+i\beta m-\alpha_3\beta\frac{1}{r}\hat k(F)],
\end{equation}
where 
\begin{equation}
\hat k(F)~=~S^{-1}_0\hat K(F)S_0~=~-i\beta\Sigma_2\frac{\partial}{\partial\theta} + i\beta\Sigma_1\frac{1}{sin\theta}(\frac{\partial}{\partial \phi} -iF).
\end{equation}
The equation (1) can be rewritten as
\begin{equation}
-i[\alpha_3\frac{\partial}{\partial r}+i\beta m-\alpha_3\beta\frac{k}{r}]\Psi(r,\theta,\phi)~=~E\Psi(r,\theta,\phi),
\end{equation}
where
\begin{equation}
\hat k(F)\Psi(r,\theta,\phi)~=~k\Psi(r,\theta,\phi).
\end{equation}
From rotational symmetry around the z-axis, the eigenvalue $\mu$ of the z-directional angular momentum is restricted to the series of values $m+\frac{1}{2}$ where $m=0, \pm 1, \pm 2, \cdots$ and the functions $\Psi(r,\theta,\phi)$ are given by
\begin{equation}
\Psi(r,\theta,\phi)~=~e^{i\mu\phi}
\left(
\begin{array}{c}
\chi_1(r,\theta)\\
\chi_2(r,\theta)\\
\end{array}
\right).
\end{equation}
The operator $\hat k(F)$ is written as a direct sum of two operators,
\begin{equation}
\hat k(F)~=~\left(
\begin{array}{cc}
\hat k_\theta(M) & 0 \\
0 & -\hat k_\theta(M)\\
\end{array}
\right),
\end{equation}
where 
\begin{equation}
\hat k_\theta (M)~=~
-i\sigma_2\frac{\partial}{\partial\theta}-\sigma_1\frac{1}{sin\theta}M
\end{equation}
and $M=\mu-F$. After a complete separation of variables
\begin{equation}
\chi_i(r,\theta)~=~f_i(r)
\left(
\begin{array}{c}
\xi_1^i(\theta)\\
\xi_2^i(\theta)\\
\end{array}
\right),
\end{equation}
we obtain the relation
\begin{equation}
\left(
\begin{array}{c}
\xi_1^2(\theta)\\
\xi_2^2(\theta)\\
\end{array}
\right)~=~\sigma_3
\left(
\begin{array}{c}
\xi_1^1(\theta)\\
\xi_2^1(\theta)\\
\end{array}
\right)
\end{equation}
from (9). By using the equation (14) the eigenvalue equations of $\hat h(F)$ and $\hat k(F)$ are simplified 
to the following ones for two-component functions,
\begin{equation}
\hat h_r(k)\left(
\begin{array}{c}
f_1(r)\\
f_2(r)\\
\end{array}
\right)~=~
E
\left(
\begin{array}{c}
f_1(r)\\
f_2(r)\\
\end{array}
\right),
\end{equation}
where 
\begin{equation}
\hat h_r (k)~=~\left(
\begin{array}{cc}
m &
-i(\frac{d}{dr}+\frac{k}{r})\\
-i(\frac{d}{dr}-\frac{k}{r}) &
-m\\
\end{array}
\right)
\end{equation}
and
\begin{equation}
\hat k_\theta(M)\left(
\begin{array}{c}
\xi_1(\theta)\\
\xi_2(\theta)\\
\end{array}\right)~=~
k 
\left(
\begin{array}{c}
\xi_1(\theta)\\
\xi_2(\theta)\\
\end{array}\right),
\end{equation}
respectively.

We now check on the self-adjointness of the Hamiltonian $\hat H(F)$ and the operator $\hat K(F)$ and construct their self-adjoint extensions. These problems are reduced to ones of the radial and the angular parts $\hat h_r(k)$ and $\hat k_\theta(M)$. The argument on $\hat h_r(k)$ is similar to one given in \cite{3,5}, and then it is not repeated here. Hereafter, we shall restrict ourselves to the operator $\hat k_\theta (M)$ on the Hilbert space defined by the inner product 
\begin{equation}
<\zeta|\xi>~=~\int_0^\pi d\theta \sum_{i=1}^2 \zeta_i^*(\theta)\xi_i(\theta).
\end{equation}
The measure is modified because of the similarity transformation (4). The domain of definition for $\hat k_\theta(M)$ is established in the following process.   

First, the space of the two-component functions satisfying the custamary condition of nonsingular wave function is taken  as a trial domain $D(\hat k_\theta(M))$ for $\hat k_\theta(M)$ and we solve the equation (17), which reads
\begin{equation}
(\frac{d}{d\theta}-\frac{1}{sin\theta}M) \xi_1(\theta)~=~
 k \xi_2(\theta),~~~~~
(-\frac{d}{d\theta}-\frac{1}{sin\theta}M) \xi_2(\theta)~=~
k \xi_1(\theta).
\end{equation}
Eliminating $\xi_2(\theta)$ yields an equation for $\xi_1(\theta)$
\begin{equation}
(\frac{d^2}{dx^2}-\frac{M(M-1)}{sin^2x}-\frac{M(M+1)}{cos^2x}+
(2k)^2) \xi_1(x)~=~0,
\end{equation}
where $x=\frac{\theta}{2}$. Applying the factorization method for solving eigenvalue problems \cite{8} to the equation (20) and substituting these solutions into (19), we can obtain the eigenvalues and eigenfunctions of (17). The results only are summarized here.
\begin{enumerate} 
\item
\begin{enumerate}
\item
For $\frac{1}{2} \leq M$, $k=\pm |n+1-F|$ where n is an integer value in the region $n\geq M+F-\frac{1}{2}$.
\item
For $-\frac{1}{2} \geq M$, $k=\pm |n+F|$ where n is an integer value in the region $n\leq M+F-\frac{1}{2}$.
\item
For $-\frac{1}{2} < M < \frac{1}{2}$, there is no solution satisfying the custamary condition.
\end{enumerate}
\item
When F is nonintegral, the wave function obtained from the eigenfunctions vanishes at the z-axis ($\theta=0$ and $\pi$). The particle, thus, does not touch the magnetic flux. 
\end{enumerate}
It is noteworthy that when the magnetic flux $F$ is nonintegral M takes a value whithin the region (c). Since it means a loss of the completeness in the angular basis around the z-axis, it is easy to see that the above-mentioned domain is insufficient.

Secondly, we try to construct the self-ajoint extensions of $\hat k_\theta(M)$ by applying the von Neumann theory of deficiency indices \cite{2}. The theory need the adjoin operator $\hat k^*_\theta(M)$ defined by the same differential operator (12). But it acts on a different domain from $D(\hat k_\theta(M))$, which is so large that the entire complex plan is included in the spectrum of $\hat k^*_\theta(M)$.  It is well-known that the definciency space of $\hat k_\theta(M)$ is generated by the normalizable eigenfunctions of the following equation
\begin{equation}
\hat k^*_\theta(M)\left(
\begin{array}{c}
\zeta_1(\theta)\\
\zeta_2(\theta)\\
\end{array}\right)~=~
\pm i 
\left(
\begin{array}{c}
\zeta_1(\theta)\\
\zeta_2(\theta)\\
\end{array}\right).
\end{equation}
In order to solve the equation (21) we make the following ansatz;
\begin{equation}
\zeta_1(z)~=~z^{\frac{M}{2}}(1-z)^{\frac{1+M}{2}}\eta_1(z),~~~
\zeta_2(z)~=~z^{\frac{1+M}{2}}(1-z)^{\frac{M}{2}}\eta_2(z)
\end{equation}
where $z=sin^2\frac{\theta}{2}$. Substituting (22) into (21) we obtain the coupled equatins,
\begin{equation}
-(\frac{1}{2}+M)\eta_1(z) +(1-z)
\frac{d}{dz}\eta_1(z) \pm i\eta_2(z)~=~0
\end{equation}
\begin{equation}
-(\frac{1}{2}+M)\eta_2(z) -z \frac{d}{dz}\eta_2(z) \pm i\eta_1(z)~=~0.
\end{equation}
Eliminating $\eta_2(z)$ yields an equation for $\eta_1(z)$ 
\begin{equation}
(1-z)z\frac{d^2}{dz^2}\eta_1(z) +[(M+\frac{1}{2})-2(1+M)z]\frac{d}{dz}\eta_1(z)-[(\frac{1}{2}+M)^2+1]\eta_1(z)=0,
\end{equation}
of which solutions can be expressed in terms of the hypergeometric function $F(\alpha,\beta,\gamma;z)$ \cite{9}.
Two normalizable eigenfunctions of (21) for each eigenvalue are found in the region $-\frac{1}{2}<M<\frac{1}{2}$. Their representations near $z=0$ are given by
\begin{equation}
\zeta_\pm^1(z)=N_1\left(
\begin{array}{c}
z^{\frac{M}{2}}(1-z)^{-\frac{M}{2}}F(1,-1,\frac{1}{2}+M;z) \\
\pm \frac{i}{\frac{1}{2}+M}z^{\frac{1+M}{2}}(1-z)^{\frac{M}{2}}
F(-\frac{1}{2}+M,\frac{3}{2}+M,\frac{3}{2}+M;z)\\
\end{array}
\right)
\end{equation}
\begin{equation}
\zeta_\pm^2(z)=N_2\left(
\begin{array}{c}
z^{\frac{1-M}{2}}(1-z)^{-\frac{M}{2}}F(\frac{3}{2}-M,-\frac{1}{2}-M,\frac{3}{2}-M;z) \\
\mp i(\frac{1}{2}-M)z^{-\frac{M}{2}}(1-z)^{\frac{M}{2}}
F(-1,1,\frac{1}{2}-M;z)\\
\end{array}
\right)
\end{equation}
where $N_i$ is a normalization constant and $\pm$ corresponds to the sign of the eigenvalue. While, the connection formula of the hypergeometric function \cite{10} leads to the representation near $z=1$,
$$\zeta^1_\pm(z)~=~N_1\times$$
\begin{equation}
\left(
\begin{array}{c}
\frac{\Gamma(\frac{1}{2}+M)\Gamma(\frac{1}{2}+M)}{\Gamma(-\frac{1}{2}+M)\Gamma(\frac{3}{2}+M)}
z^{\frac{M}{2}}(1-z)^{-\frac{M}{2}}F(1,-1,\frac{1}{2}-M;1-z)\\
\pm\frac{i}{\frac{1}{2}+M}
z^{-\frac{M}{2}}(1-z)^{\frac{1-M}{2}}F(\frac{3}{2}-M,-\frac{1}{2}-M,\frac{3}{2}-M;1-z)
\\
\end{array}
\right),
\end{equation}
$$\zeta^2_\pm(z)~=~N_2\times$$
\begin{equation}
\left(
\begin{array}{c}
z^{\frac{M}{2}}(1-z)^{\frac{1+M}{2}}F(-\frac{1}{2}+M,\frac{3}{2}+M,\frac{3}{2}+M;1-z)\\
\mp i(\frac{1}{2}-M)\frac{\Gamma(\frac{1}{2}-M)\Gamma(\frac{1}{2}-M)}{\Gamma(\frac{3}{2}-M)\Gamma(-\frac{1}{2}-M)}
z^{-\frac{M}{2}}(1-z)^{\frac{M}{2}}F(-1,1,\frac{1}{2}+M;1-z)
\\
\end{array}
\right ).
\end{equation}
The self-adjoint extensions of $\hat k_\theta(M)$ are in one-to-one correspondence with isometries of $\zeta^i_-(z)$ onto $\zeta^j_+(z)$. Since the eigenfuctions $\zeta^1_\pm(z)$ and $\zeta^2_\pm(z)$ are orthogonal each other, such isometries are given by
\begin{equation}
\zeta_-^i(z)~\rightarrow~\sum_{j=1}^2 U_{ij}\zeta_+^j(z),
\end{equation}
where $U_{ij}$ is a 2$\times$2 unitary matrix. The corresponding self-adjoint extension $\hat k^U_\theta(M)$ of $\hat k_\theta(M)$ is described as follows:
\begin{equation}
D(\hat k^U_\theta(M))=\{\xi(z) + c\sum_{i=1}^2 (\zeta_-^i(z)+\sum_{j=1}^2U_{ij} \zeta_+^j(z))|~ \xi(z)\in D(\hat k_\theta (M)), c\in {\bf C}\}.
\end{equation} 

Thirdly, the correct domain for $\hat k_\theta(M)$ is restated in terms of boundary conditions for the two-component function in the Hilbert space defined by the inner product (18). It is easy that the boundary conditions translate into ones for the wave function. Since the operator $\hat k^*_\theta(M)$ is assumed to be symmetric, the two-component function $\varphi(z)$ in the Hilbert space must satisfy the following relations
\begin{equation}
<\hat k^*_\theta(M)(\zeta_-^i+\sum_{j=1}^2U_{ij} \zeta_+^j)|\varphi>-<\zeta_-^i+\sum_{j=1}^2U_{ij} \zeta_+^j|\hat k^*_\theta(M) \varphi>=0,~i=1,2.
\end{equation}
Therefore, the admissible boundary conditions for $\varphi(z)$ are parametrized by the unitary matrix $U$ through the equations
\begin{equation}
\lim_{z\rightarrow 0}
(\zeta_-^1(z)+\sum_{j=1}^2U_{1j} \zeta_+^j(z))^\dagger\sigma_2\varphi(z)-
\lim_{z\rightarrow 1}
(\zeta_-^1(z)+\sum_{j=1}^2U_{1j} \zeta_+^j(z))^\dagger\sigma_2\varphi(z)=0,
\end{equation} 
\begin{equation}
\lim_{z\rightarrow 0}
(\zeta_-^2(z)+\sum_{j=1}^2U_{2j} \zeta_+^j(z))^\dagger\sigma_2\varphi(z)-
\lim_{z\rightarrow 1}
(\zeta_-^2(z)+\sum_{j=1}^2U_{2j} \zeta_+^j(z))^\dagger\sigma_2\varphi(z)=0.
\end{equation} 
Substituting $F(\alpha,\beta,\gamma;0)=1$ into (26), (27), (28) and (29), we easy obtain the asymmptotic behaviors of $\zeta_\pm^i(z)$. The functions $\zeta_\pm^i(z)$ are nomalizable but they have the diverging component and the vanishing componet near $z=0$ or $z=1$, and then the functions $\varphi(z)$ have the similar properties near $z=0$ or $z=1$.

Finally, the eigenvalue problem (17) is solved  under the new boundary conditions. For $M\geq \frac{1}{2}$ and $M\leq -\frac{1}{2}$, the above-mentioned results are not changed. On the other hand, for $-\frac{1}{2}\leq M \leq \frac{1}{2}$, where there is no solution satisfying the custamary condition, we can find the normalizable eigenfunctions, 
\begin{equation}
\xi_k(z)=c_1 
\left(
\begin{array}{c}
\xi^1_k(z)^u\\
\xi^1_k(z)^d\\
\end{array}
\right )
+ c_2
\left (
\begin{array}{c}
\xi^2_k(z)^u\\
\xi^2_k(z)^d\\
\end{array}
\right ),
\end{equation}  
where $c_i$ are complex values and they are fixed by the boundary conditions and the normalization condition. Their representations near $z=0$ are
$$\xi^1_k(z)^u=
z^{\frac{M}{2}}(1-z)^{-\frac{M}{2}}F(|k|,-|k|,\frac{1}{2}+M;z),$$ \begin{equation}
\xi^1_k(z)^d=\frac{k}{\frac{1}{2}+M}z^{\frac{1+M}{2}}(1-z)^{\frac{M}{2}}
F(\frac{1}{2}+M-|k|,\frac{1}{2}+M+|k|,\frac{3}{2}+M;z)
\end{equation}
and
$$\xi^2_k(z)^u=
z^{\frac{1-M}{2}}(1-z)^{-\frac{M}{2}}F(\frac{1}{2}-M+|k|,\frac{1}{2}-M-|k|,\frac{3}{2}-M;z),$$
\begin{equation}
\xi^2_k(z)^d=\frac{\frac{1}{2}-M}{k}z^{-\frac{M}{2}}(1-z)^{\frac{M}{2}}
F(-|k|,|k|,\frac{1}{2}-M;z).
\end{equation}
We obtain the following representaions near $z=1$ from the connection formula \cite{10},
$$\xi_k^1(z)^u~=~ \frac{\Gamma(\frac{1}{2}+M)\Gamma(-\frac{1}{2}-M)}{\Gamma(|k|)\Gamma(-|k|)}z^{\frac{M}{2}}(1-z)^{\frac{1+M}{2}}\times$$
$$F(\frac{1}{2}+M-|k|,\frac{1}{2}+M+|k|,\frac{3}{2}+M;1-z)+$$
$$\frac{\Gamma(\frac{1}{2}+M)\Gamma(\frac{1}{2}+M)}{\Gamma(\frac{1}{2}+M-|k|)\Gamma(\frac{1}{2}+M+|k|)}z^{\frac{M}{2}}(1-z)^{-\frac{M}{2}}\times$$
$$F(|k|,-|k|,\frac{1}{2}-M;1-z),$$
$$\xi_k^1(z)^d~=~\frac{k}{\frac{1}{2}+M}\left (
\frac{\Gamma(\frac{3}{2}+M)\Gamma(\frac{1}{2}-M)}{\Gamma(1+|k|)\Gamma(1-|k|)}z^{-\frac{M}{2}}(1-z)^{\frac{M}{2}}\times\right.$$
$$F(-|k|,|k|,\frac{1}{2}+M;1-z)+\frac{\Gamma(\frac{3}{2}+M)\Gamma(-\frac{1}{2}+M)}{\Gamma(\frac{1}{2}+M-|k|)\Gamma(\frac{1}{2}+M+|k|)}\times$$
\begin{equation}\left.
z^{-\frac{M}{2}}(1-z)^{\frac{1-M}{2}}F(\frac{1}{2}-M+|k|,\frac{1}{2}-M-|k|,\frac{3}{2}-M;1-z)\right )
\end{equation}
and
$$\xi_k^2(z)^u~=~ \frac{\Gamma(\frac{3}{2}-M)\Gamma(-\frac{1}{2}-M)}{\Gamma(\frac{1}{2}-M+|k|)\Gamma(\frac{1}{2}-M-|k|)}z^{\frac{M}{2}}(1-z)^{\frac{1+M}{2}}\times$$
$$F(\frac{1}{2}+M-|k|,\frac{1}{2}+M+|k|,\frac{3}{2}+M;1-z)+$$
$$
\frac{\Gamma(\frac{3}{2}-M)\Gamma(\frac{1}{2}+M)}{\Gamma(1-|k|)\Gamma(1+|k|)}z^{\frac{M}{2}}(1-z)^{-\frac{M}{2}}
F(|k|,-|k|,\frac{1}{2}-M;1-z)
,$$
$$\xi_k^2(z)^d~=~\frac{\frac{1}{2}-M}{k}\left (
\frac{\Gamma(\frac{1}{2}-M)\Gamma(\frac{1}{2}-M)}{\Gamma(\frac{1}{2}-M+|k|)\Gamma(\frac{1}{2}-M-|k|)}z^{-\frac{M}{2}}\times\right.$$
$$(1-z)^{\frac{M}{2}}
F(-|k|,|k|,\frac{1}{2}+M;1-z)+\frac{\Gamma(\frac{1}{2}-M)\Gamma(-\frac{1}{2}+M)}{\Gamma(-|k|)\Gamma(|k|)}\times$$
\begin{equation}\left.
z^{-\frac{M}{2}}(1-z)^{\frac{1-M}{2}}F(\frac{1}{2}-M+|k|,\frac{1}{2}-M-|k|,\frac{3}{2}-M;1-z)\right ).
\end{equation}
We should notice that on the new boundary condition there exist wave functions which do not vanish at the z-axis. 

The details of the eigenvalue $k$ and the corresponding physical phenomena will be given in \cite{7} together with the results in the presence of the Coulomb and the magnetic monopole fields.
\section*{Acknowledgements}
One of us (K.O) has benefited from conversations with M. Omote, S. Sakoda, Y. Ohnuki, S. Kamefuti and Y. Takahashi. He is grateful to them for their assistance.

To Professor Tsuneo Katayama we dedicate this paper in token of our gratitude at his retirement.

\end{document}